\documentclass[prl,twocolumn,showpacs,showkeys]{revtex4}

\usepackage{amssymb, amsmath, amsfonts, latexsym, verbatim, mathrsfs}

\usepackage{graphicx}

\begin{document}

\title{A general hidden variable model for the two-qubits system}
\thanks{One of us, R.R., acknowledges the financial support by the R. Parisi Foundation }

\author{GianCarlo Ghirardi}
\email{ghirardi@ictp.it}
\affiliation{Department of Physics, University of Trieste, the Abdus Salam ICTP, Trieste and INFN, Sezione di Trieste, Italy}
\author{Raffaele Romano}
\email{rromano@ts.infn,it}
\affiliation{Department of Physics, University of Trieste, Fondazione Parisi, Rome, Italy}


\begin{abstract}

\noindent We generalize  Bell's hidden variable model describing the singlet state of a two-qubits system by extending it to arbitrary states and observables. As in the original work, we assume a uniform, state-independent probability distribution for the hidden variables which are identified with the unit vectors of a 3-dimensional real space. By slightly modifying our model, we  provide also a minimal hidden variable description of the two-qubits system, relying on a single hidden variable. We discuss the main features and the implications of the model.
\end{abstract}

\pacs{03.65.Ta, 03.65.Ud}

\keywords{Hidden Variable Theories, Nonlocality, Contextuality}

\maketitle


{\it Introduction ---} Hidden variables theories aim at reducing  the weirdness of the quantum world relating it to incomplete knowledge. The search for such theories has been stimulated by the famous  EPR paper \cite{Einstein} suggesting that quantum mechanics does not provide a complete description of the microscopic word. The real challenge is to work out model theories in which even  non-commuting observables might be thought of as always having definite values. In spite of von Neumann's ``proof"  that such a line is unviable \cite{vNeumann}, the hidden variables program has been extensively developed, the most well-known example being represented by Bohmian Mechanics \cite{Bohm}.

While in quantum mechanics the vector $\vert \psi \rangle$ fully describes the state of a system, in the most popular versions of hidden variables theories the complete description requires additional, unaccessible variables, denoted by $\lambda$, with statistical distribution $\rho_{\psi}(\lambda)$. In these theories, an observable $Q$ always assumes a well defined value $Q_{\psi}(\lambda)$ - coinciding with one of the eigenvalues of the associated quantum operator - which in general depends on both $\vert \psi \rangle$ and $\lambda$. Obviously, one must require that the hidden variables theories one considers be predictively equivalent to quantum mechanics. Lack of knowledge of the effective values of $\lambda$ is responsible for the statistical spread of quantum measurements of $Q$ in the state $\vert \psi \rangle$ so that, while in the ordinary formulation of quantum mechanics the probabilities are non-epistemic, in these models they become epistemic, due to our ignorance concerning the hidden variables.

As already mentioned, the first consistent hidden variables model  has been presented by Bohm \cite{Bohm}; it provides a description where, contrary to the standard formulation of quantum mechanics, the particles possess well defined trajectories. It has to be stressed that hidden variables theories have played a fundamental role in leading to the identification of two basic aspects of nature: non-locality and contextuality. In fact it is well known that the critical analysis of the bohmian version of the EPR situation has led Bell to derive \cite{Bell} his celebrated inequality which implies that any theory whatsoever agreeing with the quantum statistical predictions for entangled states must exhibit nonlocal causality. Here, we will deal with nonlocal hidden variables theories for the system of 2 spin $1/2$ particles. We mention that a different hidden variables model based on a completely different approach, which explicitly resorts to classical communication, but as general as ours, has been outlined in \cite{Toner}.

{\it Bell's model ---} In his famous paper, Bell has presented a simple hidden variables model for a pair of spin $1/2$ particles which reproduces all quantum predictions concerning single or correlated measurements of spin components, but, it has to be stressed, for the exclusive case in which the system is in the singlet state,
\begin{equation}\label{singlet}
    \vert \phi_{-} \rangle = \frac{1}{\sqrt{2}} (\vert 01 \rangle - \vert 10 \rangle).
\end{equation}
 We denote the spin local observables (with spectrum $\{-1, +1\}$) of the two particles, as   $A^{(1)} = \sigma^{(1)} \cdot a$ and $B^{(2)} = \sigma^{(2)} \cdot b$, where $a$ and $b$ are $3-$dimensional, real, unit vectors representing the measurement settings independently chosen by the two parties, and $\sigma^{(i)} = (\sigma^{(i)}_x, \sigma^{(i)}_y, \sigma^{(i)}_z)$, $(i=1,2)$ are the vectors of Pauli matrices.

  In the model, the hidden variables are identified with a $3-$dimensional, real, unit vector $\lambda$ with uniform distribution over the sphere, $\rho (\lambda) = 1/4 \pi$, and the aforementioned observables assume precise values depending only on $\lambda$.

Bell gives, first of all, in terms of $\lambda$, the values taken by the observables when only one of the two systems is subjected to a measurement process:

  \begin{equation}\label{bellobs1}
    A^{(1)}(a; \lambda) = {\rm sign} (a \cdot \lambda), \quad B^{(2)}(b; \lambda) = -{\rm sign} (b \cdot \lambda),
\end{equation}

\noindent and then he introduces nonlocality by assuming that, in the case of joint measurements along $a$ and $b$, the values are:

\begin{equation}\label{bellobs}
    A^{(1)}(a, b;\lambda) = {\rm sign} (\hat{a} \cdot \lambda), \quad B^{(2)}(a,b; \lambda) = -{\rm sign} (b \cdot \lambda),
\end{equation}
where $\hat{a} = \hat{a} (a, b)$ is defined by $\hat{\theta} = \pi (1 - \cos{\theta})/2$, with $\cos{\theta} = a \cdot b$, $\cos{\hat{\theta}} = \hat{a} \cdot b$, and $\theta, \hat{\theta} \in [0, \pi]$.  Non-locality is apparent since the outcome of the measurement of $A^{(1)}$  depends explicitly on $b$. Note that nonlocality is asymmetrically distributed between the two particles. The quantum mechanical averages,
\begin{eqnarray}\label{qmcorr}
 \nonumber \langle A^{(1)} \rangle_{\phi_-} &=& \langle \phi_- \vert A^{(1)} \otimes I^{(2)} \vert \phi_- \rangle = 0, \\
  \langle B^{(2)} \rangle_{\phi_-} &=& \langle \phi_- \vert I^{(1)}\otimes B^{(2)} \vert \phi_- \rangle = 0, \\
 \nonumber \langle A^{(1)} \otimes B^{(2)} \rangle_{\phi_-} &=& \langle \phi_- \vert A^{(1)} \otimes B^{(2)} \vert \phi_- \rangle = - a \cdot b,
\end{eqnarray}
are reproduced as averages over $\lambda$, that is, one requires that $\langle Q \rangle_{\lambda} = \langle  Q \rangle_{\phi_-}$, where
\begin{equation}\label{avlambda}
    \langle Q\rangle_{\lambda} = \int Q(\lambda) \rho(\lambda) d \lambda
\end{equation}
for $Q = A^{(1)}, B^{(2)}, A^{(1)} \otimes B^{(2)}$ respectively.

Thus, it is sufficient to add to $\vert \phi_- \rangle$, the singlet state, a $3-$dimensional, real, unit vector $\lambda$ with uniform distribution, to assign a well defined value to all single and joint  spin components of the constituents of the system. The price to be paid is to introduce non-locality in the description.  The model is interesting since it describes the system of two-qubits in the singlet state, which is widely studied in quantum information, computation and control; moreover, it is very simple, since only two hidden variables are required (for instance, the two polar coordinates of $\lambda$), with uniform distribution.

A question naturally arises: Bell has limited his considerations to the singlet state and to the spin components as observables. Would it be possible to work out a similar nonlocal hidden variables model which reproduces quantum mechanics for individual and joint but factorized arbitrary observables and for arbitrary  states  of two qubits? This requires to identify the appropriate variables and to give precise rules which assign definite values to all single and joint measurements of  hermitian operators of the Hilbert spaces of the constituents. In the literature this  problem has never been considered, to our knowledge, in spite of the fact that there are various cases in which one has to deal with non maximally entangled states. Moreover,  if this can be done, what lesson can be learned out of it?

{\it The problem ---}With these motivations in mind,  we generalize  Bell's model by considering: (i) arbitrary local observables $X^{(1)}$ and $Y^{(2)}$; (ii) an arbitrary pure state $\vert \psi \rangle$ for the composite system.
As a characteristic feature of  Bell's model, we maintain a uniform distribution $\rho(\lambda)$. This means that $\vert \psi \rangle$ and $\lambda$ are independent quantities, since the knowledge of the state does not provide any information on the hidden variables. The obvious price to be paid to accomplish this task is to introduce an explicit dependence on the state in the expression (\ref{bellobs}) for the value of the observables $X^{(1)}\otimes Y^{(2)}$ which will, as before, be uniquely determined by the assignement of $\psi$ and $\lambda$
\footnote {Note that, by taking either $X^{(1)}$ or $Y^{(2)}$ equal to the identity, one recovers the local observables.}.
The model represents a special case of the general framework, where both the values of the observables and the statistical distribution of the hidden variables are state-dependent.

Finally, inspired by this model, we exhibit a minimal hidden variables model for the two-qubits system, in which the hidden variable consists of a single real parameter.

{\it Quantum averages ---} To account for (i), we consider observables of the form $X^{(1)} = \alpha_1 I ^{(1)}+ \alpha_2 \sigma^{(1)} \cdot a$ and $Y^{(2)}= \beta_1 I^{(2)} + \beta_2 \sigma^{(2)} \cdot b$, with $\alpha_1$, $\beta_1, \alpha_2$, $\beta_2 \in \texttt{R}$, and $a, b$ as before. It is always possible to consider $\alpha_2, \beta_2 \geqslant 0$ by suitably choosing $a$ and $b$, and we adopt this convention. Relations (\ref{qmcorr}) can be compactly expressed as
\begin{equation}\label{qmcorr2}
    \langle X^{(1)} \otimes Y^{(2)} \rangle_{\phi_-} = \alpha_1 \beta_1 - \alpha_2 \beta_2 \, a \cdot b;
\end{equation}
by putting $Y^{(2)} = I^{(2)}$, or $X^{(1)} = I^{(1)}$, we obtain the averages of the single, local observables, $\langle X^{(1)} \rangle_{\phi_-} = \alpha_1$, and $\langle Y^{(2)} \rangle_{\phi_-} = \beta_1$.

In order to take into account (ii) we use the Schmidt decomposition for the state $\vert\psi\rangle$,

\begin{equation}\label{genstate}
    \vert \psi \rangle = \sqrt{\mu_1} \, \vert a_1 b_1 \rangle + \sqrt{\mu_2} \, \vert a_2 b_2 \rangle,
\end{equation}

\noindent where $\mu_1 \in [1/2,1]$ without loss of generality, $\mu_1 + \mu_2 = 1$, and $\{\vert a_i \rangle, i = 1, 2\}$, $\{\vert b_i \rangle, i = 1, 2\}$ are orthonormal bases in the Hilbert spaces pertaining to the two parties. We can always write $\vert \psi \rangle = N^{(1)}U^{(1)} \otimes I^{(2)} \vert \phi_- \rangle$, where $U^{(1)}$ and $N^{(1)}$ are a unitary operator and an Hermitian operator of the Hilbert space of the first particle \footnote {In fact we can define an operator $U^{*}=U^{(1)}\otimes U^{(2)}$, where $U^{(1)}:\vert 0\rangle \rightarrow \vert a_{1}\rangle, \vert 1\rangle \rightarrow -\vert a_{2}\rangle $ and $U^{(2)}:\vert 0\rangle \rightarrow \vert b_{2}\rangle, \vert 1\rangle \rightarrow \vert b_{1}\rangle $. We then take advantage of the fact that, for any hermitian operator $\Gamma$ of the space of a particle and any maximally entangled state $\vert \Phi\rangle,$ it holds: $I^{(1)}\otimes \Gamma^{(2)}\vert \Phi\rangle= \Gamma^{T(1)}\otimes  I^{(2)}\vert \Phi\rangle$ in order to express the map we are interested in in terms of operators of the Hilbert space of only particle 1. In what follows we will use the symbol $U$ to represent the unitary operator $U^{(1)}\cdot [U^{(2)T}]^{(1)}$. }. The operator $U^{(1)}$ exists because all the maximally entangled states are locally equivalent, and $N^{(1)}$ is defined by $\vert a_i \rangle \rightarrow \sqrt{2 \mu_i} \, \vert a_i \rangle$, $i = 1, 2$.

We find convenient to write $N^{(1)} = \cos{\varphi} \, I^{(1)} + \sin{\varphi} \,\sigma^{(1)}\cdot n$, where $n$ is a the $3-$dimensional, real unit vector such that $\sigma^{(1)}\cdot n\vert a_{1}\rangle=\vert a_{1}\rangle$, and $\varphi \in [0, \pi/4]$. The angle $\varphi$ is a measure of entanglement: if $\varphi = 0$, $\vert \psi \rangle$ is a maximally entangled state; if $\varphi = \pi/4$ it is a separable state. For further reference, we notice that $N^{(1)2} = I^{(1)} + \sin{2 \varphi} \, \sigma^{(1)}\cdot  n$. From here on we will drop, for simplicity, the apices (1) and (2) from the operators which appear.

The quantum mechanical averages in the state $\vert \psi \rangle$ are given by
\begin{equation}\label{qmcorr3}
 \langle X \otimes Y \rangle_{\psi} = \langle \tilde{X} \otimes Y \rangle_{\phi_-} = \tilde{\alpha}_1 \beta_1 - \tilde{\alpha}_2 \beta_2 \, \tilde{a} \cdot b,
\end{equation}
where $\tilde{X} = N^{\prime} X^{\prime} N^{\prime} = \tilde{\alpha}_1 I + \tilde{\alpha}_2 \sigma \cdot \tilde{a}$, with $X^{\prime} = U^{\dagger} X U = \alpha_1 I + \alpha_2 \sigma \cdot a^{\prime}$, and $N^{\prime} = U^{\dagger} N U = \cos{\varphi} \, I + \sin{\varphi} \, \sigma \cdot n^{\prime}$. The vectors $a^{\prime}$ and $n^{\prime}$ are obtained by rotating $a$ and $n$ through the same orthogonal transformation, induced by $U$. It turns out that
\begin{eqnarray}
  \tilde{\alpha}_1 &=& \alpha_1 + \alpha_2 \sin{2 \varphi} \, a^{\prime} \cdot n^{\prime}, \\
  \nonumber \tilde{\alpha}_2 \tilde{a} &=& \alpha_2 \cos{2 \varphi} a^{\prime} + (\alpha_1 \sin{2 \varphi} + 2 \alpha_2 \sin^2{\varphi} \, a^{\prime} \cdot n^{\prime}) n^{\prime}.
\end{eqnarray}
From (\ref{qmcorr3}) we obtain $\langle X \rangle_{\psi} = \tilde{\alpha}_1$ and $\langle Y \rangle_{\psi} = \langle N^{\prime 2} \otimes Y \rangle_{\psi}$, more explicitly
\begin{eqnarray}
  \nonumber \langle X \rangle_{\psi} &=& \alpha_1 + \alpha_2 \sin{2 \varphi} \, a^{\prime} \cdot n^{\prime}, \\
  \langle Y \rangle_{\psi} &=& \beta_1 - \beta_2 \sin{2 \varphi} \, b \cdot n^{\prime}
\end{eqnarray}
Obviously, every hidden variables model must satisfy $\langle Q\rangle_{\lambda} = \langle Q \rangle_{\psi}$ for $Q = X, Y, X \otimes Y$. Now we impose these conditions to our model.

{\it Generalizing Bell's model ---} We assume that $\rho(\lambda)$ is the uniform distribution, and write, completely in general:
\begin{eqnarray}\label{bellobs2}
    \nonumber X_{\psi}(a,b,\lambda) &=& \alpha_1 + \alpha_2 F_{\psi}(a, \hat{a}, \lambda),  \\
    Y_{\psi}(b,\lambda) &=& \beta_1 + \beta_2 G_{\psi}(b, \lambda),
\end{eqnarray}
where the two functions $F_{\psi}$ and $G_{\psi}$ assume values in $\{-1, 1\}$ in order to reproduce the spectra of $X$ and $Y$, $\{ \alpha_1 \pm \alpha_2\}$ and $\{ \beta_1 \pm \beta_2\}$ respectively. As before, $\hat{a} = \hat{a} (a, b)$ is a function to be determined.


We define
\begin{equation}\label{deff}
    F_{\psi} (a, \hat{a}, \lambda) = \left\{
                                    \begin{array}{ll}
                                      +1, & \hbox{\rm if $\hat{a} \cdot \lambda \geqslant \cos{\xi}$,} \\
                                      -1, & \hbox{\rm if $\hat{a} \cdot \lambda < \cos{\xi}$,}
                                    \end{array}
                                  \right.
\end{equation}
where $\xi = \xi(a, \psi) \in [0, \pi]$ is fixed by $\langle X \rangle_{\lambda} = \langle X \rangle_{\psi}$, leading to $\cos{\xi} = -\sin{2 \varphi} \, a \cdot n$ without any constraint on $\hat{a}$. Similarly,
\begin{equation}\label{defg}
    G_{\psi} (b, \lambda) = \left\{
                                    \begin{array}{ll}
                                      +1, & \hbox{\rm if $b \cdot \lambda < \cos{\chi}$,} \\
                                      -1, & \hbox{\rm if $b \cdot \lambda \geqslant \cos{\chi}$,}
                                    \end{array}
                                  \right.
\end{equation}
with $\chi = \chi(a, \psi) \in [0, \pi]$, and $\cos{\chi} = -\sin{2 \varphi} \, b \cdot n^{\prime}$, as required  by  $\langle Y \rangle_{\lambda} = \langle Y \rangle_{\psi}$. The correlations on joint measurements must satisfy $\langle X \otimes Y \rangle_{\lambda} = \langle X \otimes Y \rangle_{\psi}$, that can be fulfilled if and only if
\begin{equation}\label{corrfg}
    \langle F_{\psi} G_{\psi} \rangle_{\lambda} = - r \cdot b, \quad r = \cos{2 \varphi} \, a^{\prime} + 2 a^{\prime} \cdot n^{\prime} \sin^2{\varphi} \, n^{\prime}.
\end{equation}
To show that this relation can always be satisfied we define
\begin{eqnarray}\label{fgmaxmin}
    \nonumber \langle F_{\psi} G_{\psi} \rangle_{Min} &=& \min_{\hat{a}} \langle F_{\psi} G_{\psi} \rangle_{\lambda}, \quad \\
    \langle F_{\psi} G_{\psi} \rangle_{Max} &=& \max_{\hat{a}} \langle F_{\psi} G_{\psi} \rangle_{\lambda},
\end{eqnarray}
and observe that, for fixed $a^{\prime}$, $b$, $n^{\prime}$ and $\varphi$, $\langle F_{\psi} G_{\psi} \rangle_{\lambda}$ assumes any value in the interval $[\langle F_{\psi} G_{\psi} \rangle_{Min}, \langle F_{\psi} G_{\psi} \rangle_{Max}]$. We shall shortly show that, under the same conditions,
\begin{equation}\label{reqf}
    \langle F_{\psi} G_{\psi} \rangle_{Min} \leqslant - r \cdot b \leqslant \langle F_{\psi} G_{\psi} \rangle_{Max}.
\end{equation}
It is possible to prove that $\langle F_{\psi} G_{\psi} \rangle_{\lambda}$ is minimal when $\hat{a} = b$, and maximal when $\hat{a} = -b$. Therefore, the explicit forms of (\ref{fgmaxmin}) are given by
\begin{eqnarray}\label{fgmaxminexpl}
    \nonumber \langle F_{\psi} G_{\psi} \rangle_{Min} &=& \vert \cos{\xi} - \cos{\chi} \vert - 1, \quad \\
    \langle F_{\psi} G_{\psi} \rangle_{Max} &=& 1 - \vert \cos{\xi} + \cos{\chi} \vert.
\end{eqnarray}
{\it Proof ---} We find convenient to define $\tau, \sigma, \omega \in [0, \pi]$ such that $a \cdot n = \cos{\tau}$, $b \cdot n^{\prime} = \cos{\sigma}$, and $a^{\prime} \cdot b = \cos{\omega}$. Therefore, (\ref{fgmaxminexpl}) can be rewritten as
\begin{eqnarray}\label{fgmaxminexp2}
    \nonumber \langle F_{\psi} G_{\psi} \rangle_{Min} &=& \sin{2 \varphi} \, \vert  \cos{\sigma} - \cos{\tau} \vert - 1, \quad \\
    \langle F_{\psi} G_{\psi} \rangle_{Max} &=& 1 - \sin{2 \varphi} \, \vert \cos{\sigma} + \cos{\tau} \vert,
\end{eqnarray}
and $r \cdot b = \cos{2 \varphi} \, \cos{\omega} + 2 \sin^2{\varphi} \, \cos{\tau} \, \cos{\sigma}$. Now, any triple of unit vectors must satisfy $\omega \leqslant \min [\tau + \sigma, 2 \pi - (\tau + \sigma)]$ and $\omega \geqslant \vert \tau - \sigma \vert$. From these relations it follows that necessarily
\begin{equation}\label{relangles}
    \vert \cos{\omega} - \cos{\tau} \, \cos{\sigma} \vert \leqslant \sin{\tau} \, \sin{\sigma},
\end{equation}
 which in turn implies (\ref{reqf}). Therefore, there exists a vector $\hat{a}$ such that (\ref{corrfg}) holds, and all the quantum mechanical correlations are reproduced. In general, $\hat{a}$ depends on both $a$ and $b$, a signature on non-locality, and it could also depend on the state, such that $\hat{a} = \hat{a}_{\psi} (a, b)$. The explicit expression of $\hat{a}_{\psi} (a, b)$ is rather lengthy and not relevant here. QED.

The Bell model is reproduced when $\vert \psi \rangle = \vert \phi_- \rangle$, $\alpha_1 = \beta_1 = 0$, and $\alpha_2 = \beta_2 = 1$. In fact, in this case $\xi = \chi = \pi/2$, and $F_{\psi} = {\rm sign} (\hat{a} \cdot \lambda)$, $G_{\psi} = - {\rm sign} (b \cdot \lambda)$.

{\it Time evolution ---} The dynamics of the system is determined by the Hamiltonian operator $H$ through the unitary propagator $V_t = e^{- i H t}$, which generates $\vert \psi \rangle \rightarrow \vert \psi_t \rangle = V_t \vert \psi \rangle$. This evolution produces time-dependent quantities $\varphi_t$, $n_t$ and $U_t$, whose explicit forms are not relevant here. Consequently, the time dependence determines the evolution of the values of the observables  through the angles $\xi_t$ and $\chi_t$. Following the previous procedure one reproduces all the quantum correlations simply by suitably defining $\hat{a}_t$. Notice that, in this model, the hidden variables $\lambda$ do not depend on time, and they are always uniformly distributed.

{\it Minimal hidden variables model ---} We modify the previous model in order to produce a model based on a single hidden variable. This is obtained by requiring that $\lambda$ belongs to a unit circle. For instance, it is sufficient to choose $\lambda_z = 0$, but there are different, equivalent choices. The assignments (\ref{bellobs2}), (\ref{deff}), and (\ref{defg}) are left unchanged, and the requirements $\langle X \rangle_{\lambda} = \langle X \rangle_{\psi}$ and $\langle Y \rangle_{\lambda} = \langle Y \rangle_{\psi}$ produce $\xi = \pi (1 + \sin{2 \varphi} \, a^{\prime} \cdot n^{\prime})/2$ and $\chi = \pi (1 + \sin{2 \varphi} \, b \cdot n^{\prime})/2$, respectively. For the joint correlations we compute
\begin{equation}\label{fgminmod}
    \langle F_{\psi} G_{\psi} \rangle_{\lambda} = \frac{2}{\pi} \min\{\xi + \chi, \hat{\theta}\} - 1.
\end{equation}
 Assuming that $\hat{\theta} \leqslant \xi + \chi$, (\ref{fgminmod}) is consistent with the quantum mechanical predictions if and only if
\begin{equation}\label{condmod}
    \hat{\theta} = \frac{\pi}{2} (1 - r \cdot b)
\end{equation}
with $r$ defined in (\ref{corrfg}). We notice that (\ref{condmod}) is compatible with our assumption $\hat{\theta} \leqslant \xi + \chi$ as a consequence of (\ref{relangles}). If $\hat{\theta} > \xi + \chi$, we cannot reproduce the required correlations in general, but only for a restricted class of observables. Accordingly, (\ref{condmod}) represents the appropriate general expression of $\hat{\theta}$.

Since $r$ depends on $a$, $b$ and $\vert \psi \rangle$, it follows that $\hat{a} = \hat{a}_{\psi} (a, b)$. From a direct inspection of (\ref{corrfg}), we observe that it is possible to have a consistent local description for arbitrary observables only if $\cos{2 \varphi} = 0$, that is $\varphi = \pi / 4$. Therefore, only separable states admit a local hidden variable model.

{\it Contextuality ---} The so-called (by D. Mermin \cite{Mermin}) Bell-Kochen-Specker Theorem implies that the deterministic, nonlocal, hidden variables models discussed in this paper, since they reproduce the quantum expectation values for a system whose Hilbert space is 4-dimensional, must exhibit a contextual character. This means that it is impossible to satisfy the (natural) requirement that if $A,B,C,...$ is a mutually commuting subset of observables satisfying a functional identity:
\begin {equation}
f(A,B,C,...)=0,
\end {equation}

\noindent then the values assigned to them $A_{\psi}(\lambda),B_{\psi}(\lambda),C_{\psi}(\lambda),....$ in an individual system, must also satisfy:

\begin {equation}
f(A_{\psi}(\lambda),B_{\psi}(\lambda),C_{\psi}(\lambda),...)=0.
\end {equation}

The most direct way to see that our model (as well as the one by Bell) exhibits contextuality is obtained by comparing the joint measurements with the single ones. In particular, it follows directly by the very choice of the values of joint and individual measurements that:

\begin {equation}
(A^{(1)}\otimes I^{(2)}\cdot I^{(1)}\otimes B^{(2)})_{\psi}(\lambda)\neq A^{(1)}_{\psi}(\lambda)\cdot B^{(2)}_{\psi}(\lambda).
\end {equation}

As is well known the standard and simple way to prove contextuality for a system like the one under consideration has been worked out by Peres \cite{Peres} by considering an appropriate array of 9 operators. His argument obviously holds also for our model.

{\it A lesson}: We have shown that one can easily generalize Bell's model to cover the case of arbitrary quantum states and of arbitrary factorized two particles operators. The most important lesson which stems from the previous analysis is that, since the model has a basically contextual nature, if one considers the whole class of the hermitian operators in the four dimensional Hilbert space, no specification has been given of the values that must be attributed to them. Typically, it turns out to be illegitimate even to claim that to evaluate $(A^{(1)}\otimes B^{(2)}+C^{(1)}\otimes D^{(2)})_{\psi}(\lambda)$ one can simply sum the precise values which the model attributes to $A^{(1)}\otimes B^{(2)}$ and $C^{(1)}\otimes D^{(2)}$. In other terms, just as Bell has been compelled to attribute to  joint measurements a value differing from the product of the values of the corresponding single measurements

\begin {equation}
-sign (\hat{a}\cdot\lambda)sign (b\cdot \lambda)\neq -sign (a\cdot\lambda)sign ( b\cdot \lambda),
\end {equation}
 \noindent the same holds for our model and one should perform an analogous analysis and work out the appropriate rules attributing precise values to all operators of the 4-dimensional Hilbert space of the conposite system.

We have not been able to prove in general that this is possible for the system under investigation. We believe that in principle this line is viable, but requires a remarkable effort.


\end{document}